\documentclass[review]{elsarticle}

\usepackage{lineno,hyperref}
\usepackage{amsmath,amsfonts,graphicx,framed}
\usepackage{comment}
\modulolinenumbers[5]

\journal{Journal of Mathematical Analysis and Applications}

\begin{document}

\newcommand{\X}{\text{X}}
\newcommand{\mc}{\mathcal}
\newcommand{\di}{\,\mathrm d}
\newcommand{\pd}{\,\partial}
\newcommand{\re}{\mathbb{R}}

\begin{frontmatter}

\title{Symmetries and conservation laws of the Euler equations in Lagrangian coordinates}

\author{Ravi Shankar}
\address{Department of Mathematics and Statistics, California State University, Chico, CA 95929, USA\\rshankar@mail.csuchico.edu\\After July 1, 2016:\\Department of Mathematics, University of Washington, Seattle, WA 98195, USA\\shankarr@uw.edu}

\begin{abstract}
We consider the Euler equations of incompressible inviscid fluid dynamics.  We discuss a variational formulation of the governing equations in Lagrangian coordinates.  We compute variational symmetries of the action functional and generate corresponding conservation laws in Lagrangian coordinates.  We clarify and demonstrate relationships between symmetries and the classical balance laws of energy, linear momentum, center of mass, angular momentum, and the statement of vorticity advection.  Using a newly obtained scaling symmetry, we obtain a new conservation law for the Euler equations in Lagrangian coordinates in n-dimensional space.  The resulting integral balance relates the total kinetic energy to a new integral quantity defined in Lagrangian coordinates.  This relationship implies an inequality which describes the radial deformation of the fluid, and shows the non-existence of time-periodic solutions with nonzero, finite energy.
\end{abstract}


\end{frontmatter}

\linenumbers

\textbf{Note:} Results modified on June 13, 2016.

\section{Introduction}
This study concerns the Euler equations of incompressible inviscid fluid dynamics.  The Euler equations provide an accurate representation of a variety of inviscid fluid flows and are used in numerous practical situations.  Mathematical interest in the Euler equations includes the open problems of classifying nonlinear blow-up, long-time behavior, and stability of solutions for various types of initial conditions.  Conservation of energy plays an important role in controlling blow-up; for example, solutions of infinite energy have been shown to blow-up in a finite amount of time \cite{Constantin2007}.

The goal of this study is to use variational symmetries of the Euler equations in Lagrangian coordinates to find conservation laws.  
Noether's First Theorem \cite{Noether} relates variational symmetries to conservation laws of a corresponding differential system. However, the Euler equations in Eulerian variables do not admit a variational formulation. 
Point symmetries of the Euler equations in Eulerian variables were obtained in \cite{Buchnev}; see also \cite{Olver1982}. There are many works on the conservation laws of the Euler equations; see \cite{Ibragimov1985} and references therein; see also \cite{Anco2003,Anco09,Anco10} for conservation laws in \textit{n}-space computed using the direct method.  Conservation laws were obtained for the constrained Euler equations in \cite{RosenhausShankar}, and for the Euler equations in vorticity formulation \cite{Oberlack2014}. 
On the other hand, the Euler equations in Lagrangian variables (see \cite{Majda,Frisch2014} 
for background) do admit a variational formulation.  Caviglia and Morro \cite{Caviglia1987} computed symmetries and conservation laws using a variational formulation for the compressible Euler equations in Lagrangian coordinates.  However, they later \cite{Caviglia1989} computed conservation laws for the incompressible Euler equations without using symmetries or a variational formulation.  We show that a simple modification to their compressible flow Lagrangian in \cite{Caviglia1987} allows for a variational formulation of incompressible flow.

In this paper we study a variational formulation of the incompressible Euler equations in Lagrangian coordinates.  
We obtain a new scaling symmetry of the action functional.  
This scaling symmetry leads via Noether's theorem to a new conservation law 
in $n$-space.  The resulting conserved quantity relates the total energy integral, a quantity defined in Eulerian variables, to a new integral quantity defined exclusively in Lagrangian variables.  This relationship allows for the quantification of the radial deformation of the fluid, 
and rules out the existence of time-periodic solutions with nonzero, finite energy.  

\section{Variational Framework}
We first introduce Lagrangian coordinates, discuss their physical interpretation, and detail explicitly the transformation of the Euler equations to Lagrangian coordinates from Eulerian variables. Then, we demonstrate the variational formulation behind the Euler equations in Lagrangian coordinates using a Lagrange multiplier (namely the pressure) of the incompressibility constraint.
\subsection{Lagrangian Coordinates}
The continuity and Euler equations of incompressible ideal fluid dynamics are as follows \cite{Landau}:
\begin{gather}
\label{mass}
\vec \nabla_x \cdot \vec u=0,\\
\label{euler}
(\partial_t +\vec u \cdot \vec \nabla_x)\vec u=-\vec \nabla_xp ,
\end{gather}
where 
$\vec u=(u^1,\dots,u^n)$ is the velocity vector, 
$\vec \nabla_x = (\partial_{x^1},\dots,\partial_{x^n})$ is the gradient, and $p=\bar{p}/\rho$ is the pressure $\bar{p}$ divided by the constant density $\rho$; we henceforth refer to $p$ as the pressure.  The velocity vector $\vec u$ and pressure $p$ are functions of 
$(\vec x,t)=(x^1,\dots,x^n,t)$.  The first condition (\ref{mass}) gives the local conservation of  mass and specifies that the fluid be incompressible.  The second states the momentum balance within a small parcel of fluid.  
The equations are considered on $n+1$ dimensional spacetime, where $n\ge 2$ is $2$ or $3$ in most applications.

The equations for ideal incompressible fluid dynamics can be recast 
using the ``Lagrangian map" 
$\vec x(\vec a,t)=(x^1,\dots,x^n)$ \cite{Lagrange,Cauchy}.  This 
smooth 
function gives the position 
$\vec x$ of a particular fluid particle given by ``label vector" $\vec a=(a^1,\dots,a^n)$ at time $t$.  At the initial time $t = 0$, the positions of the fluid particles coincide with their ``Lagrangian coordinates":  
$\vec x(\vec a,0)=\vec a$.  As time progresses, the positions of the particles deviate from their initial positions by the action of the fluid velocity 
$\vec u(\vec x,t)$.  This relationship between position and velocity can be quantified by a kinematic system of equations: 
$\partial_t\vec x(\vec a,t)=\vec u(\vec x,t)$, where $\partial_t$ denotes partial differentiation with respect to variable $t$.  For small times $t\to0$, 
Taylor's theorem gives
\begin{align}\label{t0}
\begin{split}
\vec x(\vec a,t)&\approx\vec x(\vec a,0)+t\,\,\partial_t\vec x(\vec a,0)+O(t^2),\\
&=\vec a+t\,\vec u(\vec x(\vec a,0),0)+O(t^2)\\
&=\vec a+t\,\vec u(\vec a,0)+O(t^2).
\end{split}
\end{align}

In this moving frame that follows the fluid particles, governing equations (\ref{mass}), (\ref{euler}) take the following form \cite{Majda,Frisch2014}:
\begin{align}
\label{massL}
\det\left(\frac{\partial x^i}{\partial a^j}\right)&\equiv\text{det}\left(x^i_{j}\right)=1,
\\ 
\label{eulerL}
\frac{\partial^2 x^i}{\partial t^2}&\equiv x^i_{tt}=-\frac{\partial p}{\partial a^j}\frac{\partial a^j}{\partial x^i}\equiv- p_{j}\,\tilde{x}^{ji},\hspace{10 mm} i = 1,\dots,n.
\end{align}
It is convenient to introduce some notation.  Subscripts denote partial derivatives with respect to the fluid labels $a^i$ ($i = 1,\dots,n$) or time $t$: $f_{j}\equiv\frac{\partial f}{\partial a^j}$ and $f_{t}\equiv\frac{\partial f}{\partial t}$ 
(superscript indices denote vector or matrix components).   Here, $\tilde{x}^{ji}\equiv\frac{\partial a^j}{\partial x^i}$ is the inverse of the Jacobian matrix $x^i_{j}$ such that 
$\tilde{x}^{ji}x^i_{k}=\frac{\partial a^j}{\partial x^i}\frac{\partial x^i}{\partial a^k}=\delta^{jk}$ (identity matrix).  We assume summation over repeated indices.  

Equation (\ref{massL}) can be verified to be equivalent to (\ref{mass}) using Jacobi's formula: 
\begin{align}\label{jac}
d\det A=\det A\,\,\text{tr}(A^{-1}dA),
\end{align}
where $dA$ is the infinitesimal change in matrix $A$, and $\text{tr}(A^{ij})=A^{ii}$ is the trace of $A$.  In particular, time differentiating \eqref{massL} gives
\begin{align}\label{det_t}
\partial_t \text{det}(x^i_{j})=\text{det}(x^i_{j})\text{tr}\left(\tilde{x}^{ij}x^j_{kt}\right)=\tilde{x}^{ij}x^j_{it}=\frac{\partial a^i}{\partial x^j}u^j_{i}=\frac{\partial u^j}{\partial x^j}.
\end{align}
This expression is zero upon differentiating (\ref{massL}), which recovers ($\ref{mass}$).  
That the Jacobian is unity follows from its constancy and  the initial condition $\vec x(\vec a,0)=\vec a$.  The left side of equation (\ref{euler}) can be recovered from that of (\ref{eulerL}) by using the definition of the velocity field and the chain rule for total time derivatives: 
\begin{align}\label{xtt}
x^i_{tt}=\frac{\mathrm{d}}{\mathrm{d}t}u^i(\vec x(\vec a,t),t)=u^i_{t}+x^j_{t}\,\partial_{x^j}u^i= u^i_{t}+u^j\partial_{x^j}u^i.
\end{align}

\subsection{Extremals of the Action Functional}
We show that Eqs. (\ref{massL}) and (\ref{eulerL}) are the variational Euler-Lagrange equations for an ``action" functional $\mathcal{S}$ given in the following form:
\begin{equation}\label{action}
\mathcal{S}[\vec x,p]=\int_R\left[L\left(\partial_t\vec x\right)+p\,G\left(\vec \nabla\vec x\right)\right]\di t\di^na\equiv\int_R\mathcal{L}\left(\partial_t\vec x,\vec\nabla\vec x,p\right)\di t\di^n a.
\end{equation}
Here, $L(\partial_t\vec x)=\frac{1}{2}|\partial_t\vec x|^2=\frac{1}{2}x^i_{t}x^i_{t}$ is the kinetic energy density, and $p(\vec a,t)$ functions as a Lagrange multiplier (see \cite{Hilbert}) for non-holonomic constraint function $G\left(\vec \nabla\vec x\right)=\text{det}\left(x^i_{j}\right)-1$ that accounts for the flow incompressibility (in this context, $\vec\nabla$ denotes a ``Lagrangian gradient" $\vec\nabla=(\partial_{a^1},\dots,\partial_{a^n})$).  The integration is taken over a region $R$ of $n+1$ dimensional spacetime. 
In what follows, we let $D_i$ denote total differentiation with respect to the $i$th variable.

A variational formulation has been known for incompressible flow since Lagrange \cite{Lagrange}, but the action in that case does not involve the pressure as an explicit dynamical variable.  Instead, the action is taken to be
\begin{align}\label{S}
\mc S[\vec x]=\int_R L(\partial_t\vec x)\di t\di^na
\end{align}
subject to a local incompressibility constraint $G(\vec\nabla\vec x)=0$ for all $(\vec x,t)\in R$; the pressure $p$ arises indirectly once an extremum of the action is sought.  While these two formulations yield equivalent solutions to the variational problem, we find that performing computations with \eqref{action} is more transparent.

Lagrangian $\mathcal{L}$ in \eqref{action} is a slight modification of the function used in \cite{Caviglia1987}: \\$\mathcal{L}(\vec x,\rho)=\rho_0L-\rho_0E(\rho)+\bar{p}(\rho)(\det(x^i_j)-\rho_0/\rho)$, where $E(\rho)$ is the internal energy, $\rho(\vec a,t)$ is the density, $\rho_0(\vec a)=\rho(\vec a,0)$ is the initial density, and $\bar{p}=\bar{p}(\rho)$ is the pressure.  This is a Lagrangian for the compressible and isentropic Euler equations in Lagrangian coordinates.  For these equations, the density $\rho$ is a dynamical variable and the pressure $\bar{p}=\bar{p}(\rho)$ is a known function of $\rho$.  This is in contrast to the incompressible equations we study, for which $\rho=\rho_0$ is a constant, and $\bar{p}(\vec a,t)$ is a dynamical variable.  One can obtain our Lagrangian in (\ref{action}) from that in \cite{Caviglia1987} by dividing by the constant $\rho=\rho_0$, discarding the internal energy constant $E$, and replacing $\rho$ with $p$ as the dynamical variable that $\mathcal{L}$ depends on.  Note that the same authors in the later paper \cite{Caviglia1989} claimed there was no variational formulation for incompressible flow.

For Eulerian system \eqref{mass}-\eqref{euler}, there is no variational formulation.  However, one can be found for the related system of compressible flow; see \cite{kambe} for this formulation as well as another for irrotational compressible flow, where velocity $\vec u=\vec\nabla\phi$ is restricted to a potential representation.

We seek extremals $\vec x$ and $p$ of the action functional $\mathcal{S}$.  To this end, we take $\vec x\to\vec x+\epsilon\,\vec\eta(\vec a,t)$ and $p\to p+\epsilon\,\psi(\vec a,t)$ for $\epsilon\to0$.  
Arbitrary functions $\vec\eta$ and $\psi$ are assumed to vanish on $\partial R$. 
The first variation $\delta \mathcal{S}$ of action $\mathcal{S}$ is:
\begin{align}\label{1var1}
\begin{split}
\delta \mathcal{S}&=D_\epsilon\,\mathcal{S}[\vec x+\epsilon\vec\eta,p+\epsilon\psi]\bigr\rvert_{\epsilon=0}\\
&=\int_R\left[\eta^i_{t}L_{x^i_{t}}+\psi G+p\,\text{det}\left(\vec\nabla\vec x\right)\tilde{x}^{ij}\eta^j_{i}\right]\,\di t\di^na.
\end{split}
\end{align}
We computed the last term using \eqref{jac}: 
\begin{align*}
D_\epsilon G\left(x^i_{j}+\epsilon\eta^i_{j}\right)\bigr|_{\epsilon=0}&=D_\epsilon\text{det}\left(x^i_{j}+\epsilon\eta^i_{j}\right)\bigr|_{\epsilon=0}
=\text{det}\left(\vec\nabla\vec x\right)\text{tr}\left(\tilde{x}^{ij}\eta^j_{k}\right)\\
&=\text{det}\left(\vec\nabla\vec x\right)\tilde{x}^{ij}\eta^j_{i}
\end{align*}
Since $\psi(\vec a,t)$ is an arbitrary function, it follows from (\ref{1var1}) that \\$G(x^i_{j})=\text{det}(x^i_{j})-1=0$, which recovers incompressibility condition (\ref{massL}).  Letting $p\tilde{x}^{ij}\eta^j_{i}=D_i(p\tilde{x}^{ij}\eta^j)-p\eta^j\tilde{x}^{ij}_{i}-p_{i}\tilde{x}^{ij}\eta^j$ and $\eta^i_{t}L_{x^i_{t}}=D_t(\eta^iL_{x^i_{t}})-\eta^iD_tL_{x^i_{t}}$, Eq. (\ref{1var1}) becomes:
\begin{align}\label{1var2}
\begin{split}
\delta\mathcal{S}&=\int_R\left[D_t(\eta^ix^i_{t})+D_i(p\tilde{x}^{ij}\eta^j)\right]\di t\di^na\\
&-\int_R\left(x^j_{tt}+p_{i}\tilde{x}^{ij}+p\tilde{x}^{ij}_{i}\right)\eta^j\di t\di^na.
\end{split}
\end{align}
The first integral contributes via Gauss' Theorem to surface terms that vanish on $\partial R$. 
Hence, for the first variation $\delta\mathcal{S}$ in (\ref{1var2}) to vanish for arbitrary $\vec\eta(\vec a,t)$, it follows that $$x^j_{tt}+p_{i}\tilde{x}^{ij}+p\tilde x^{ij}_i=0$$.
To recover \eqref{eulerL}, we show that
\begin{align}\label{xiji}
\tilde x^{ij}_i=0,\qquad j=1,\dots,n,
\end{align}
when \eqref{massL} holds.  Since $\tilde{x}^{ij}\,x^j_{k}=\delta^{ik}$, differentiation and summation with respect to $i$ gives $\tilde{x}^{ij}_{i}\,x^j_{k}+\tilde{x}^{ij}\,x^j_{ik}=0$, from which we see that $\tilde{x}^{ik}_{i}=-\tilde{x}^{ij}\,x^j_{iq}\tilde{x}^{qk}$, which re-expresses the last term in (\ref{1var2}).  But the $q$th partial derivative of $\text{det}(x^i_{j})=1$ is $\partial_q\,\text{det}(x^i_{j})=0=\text{tr}\left(\tilde{x}^{ij}x^j_{kq}\right)=\tilde{x}^{ij}\,x^j_{iq}$, so $\tilde{x}^{ik}_{i}=-(\tilde{x}^{ij}x^j_{iq})\tilde{x}^{qk}=0$.  

\section{Symmetries of the Action}
In this section, we find all 
(variational) point symmetries of the action functional $\mathcal{S}$ given in (\ref{action}).  In the following section, we find local conservation laws corresponding to these variational symmetries.

\subsection{Variational symmetries}
By a variational point symmetry of the action (\ref{action}), we mean an infinitesimal Lie group transformation 
 depending on $(t,\vec a,\vec x,p)$ (no derivatives)
\begin{align*}
t\to t^*=t+\epsilon\,\xi^t(t,\vec a,\vec x,p),\\
\vec a\to\vec a^*=\vec a+\epsilon\,\vec\xi(t,\vec a,\vec x,p),\\
\vec x\to\vec x^*=\vec x+\epsilon\,\vec\eta(t,\vec a,\vec x,p),\\
p\to p^*=p+\epsilon\,\psi(t,\vec a,\vec x,p)
\end{align*}
that leaves the action integral (\ref{action}) unchanged 
for every region $R$ of spacetime.  An equivalent requirement for a variational symmetry is for the transformation to leave the action differential $\mathrm{d}\mathcal{S}=\mathcal{L}\di t\di^na$ invariant throughout the integration region, such that: 
\begin{equation}\label{invar0}
\mathcal{L}(\partial_{t^*}\vec x^*,\vec\nabla^*\vec x^*,p^*)\mathrm{d}t^*\mathrm{d}^na^*=\mathcal{L}(\partial_{t}\vec x,\vec\nabla\vec x,p)\di t\di^na.  
\end{equation}
A more general definition of variational symmetry allows for $\mc L$ to change by a total divergence (see e.g. \cite{Olver}):
\begin{equation}\label{invar}
\mathcal{L}(\partial_{t^*}\vec x^*,\vec\nabla^*\vec x^*,p^*)\,\mathrm{d}t^*\mathrm{d}^na^*=\left(\mathcal{L}(\partial_{t}\vec x,\vec\nabla\vec x,p)+\epsilon\,D_\mu N^\mu\right)\di t\di^na,
\end{equation}
where $D_\mu N^\mu=D_tN^t+D_iN^i$ is a total divergence. 

Extended to act on derivatives of $\vec x$, this point transformation has infinitesimal generator $\text{X}=\xi^\mu\partial_\mu+\eta^i\partial_{x^i}+\psi\partial_p+\zeta^{i\mu}\partial_{x^i_{\mu}}$, where 
\begin{align}\label{zeta}
\zeta^{i\mu}=D_\mu\eta^i-x^i_{\nu}D_\mu\xi^\nu,\qquad i=1,\dots,n,\qquad\mu=t,1,\dots,n.
\end{align}
The Lagrangian changes to first order according to $\mathcal{L}^*=\mathcal{L}+\epsilon \text{X}\mathcal{L}$.
The volume element $\di t\di^na$ changes to first order in $\epsilon$ under the infinitesimal transformation according to $\mathrm{d}t^*\mathrm{d}^na^*=\left[1+\epsilon(D_\mu\xi^\mu)\right]\di t\di^na$, which can be shown by direct expansion in $\epsilon$ or using 
\eqref{jac}.   Expressing these changes in (\ref{invar}), we obtain:
\begin{equation}\label{invarX}
\text{X}\mathcal{L}+\mathcal{L}\,D_\mu\xi^\mu=D_\mu N^\mu.
\end{equation}
Finding the variational symmetries of the action integral (\ref{action}) means solving invariance condition (\ref{invarX}) for the coefficients $\xi^t,\,\xi^i,\,\eta^i,\text{ and }\psi$ of infinitesimal generator X 
(as well as the unknown fluxes $N^t,N^i$ of the transformation).  


\subsection{Solution}
The technique for solving \eqref{invarX} is to equate all coefficients of derivatives of $x^i$ and $p$ to zero, since the point symmetry coefficients $\xi^\mu,\eta^i,$ and $\psi$ only depend on $(t,\vec a,\vec x,p)$.  


Since the left hand side of \eqref{invarX} depends on derivatives of order at most 1 (i.e. no second or higher derivatives), terms on the right hand side depending on higher order derivatives must be independent of the left hand side, so such terms must vanish identically in the expression $D_\mu N^\mu$.  We conclude that $N^\mu=N^\mu(t,\vec a,\vec x,p),\mu=t,1,\dots,n$, up to terms that do not influence $\X$.

We denote $\mathcal{L}=L+p(J-1)$, $J=\det(x^i_j)$.  Separating terms in \eqref{invarX} that are at most linear in $x^i_j$ from those that are at least quadratic in $x^i_j$ yields the system
\begin{align}
\label{sa1}
\X L-\psi+(L-p)D_\mu\xi^\mu&=D_\mu N^\mu,\\
\label{sa2}
p\X J+\psi J+p JD_\mu\xi^\mu&=0.
\end{align}
From \eqref{jac}, $\X J=J\tilde x^{ij}\zeta^{ji}$.  Solving \eqref{sa2} gives
\begin{align}
\psi=-p(\xi^t_t+\eta^j_{x^j}),\qquad \eta^i=\eta^i(t,\vec x),\qquad \xi^t=\xi^t(t),\qquad \xi^i=\xi^i(t,\vec a).
\end{align}
From the $x^i_tx^i_j$ term on the left hand side of \eqref{sa1} and the $x^j_i$ and $p_\mu$ terms on the right hand side,
\begin{align}\label{time}
N^t_p=N^i_p=0,\qquad\xi^i=\xi^i(\vec a),\qquad N^i=N^i(t,\vec a),\qquad i=1,\dots,n.
\end{align}
Collecting the remaining terms of zeroth, first, and second order in powers of derivatives gives the system
\begin{align}
\label{sb1}
p(\eta^j_{x^j}-\xi^j_j)=N^t_t+N^i_i,\\
\label{sb2}
x^i_t\eta^i_t=x^i_tN^t_{x^i},\\
\label{sb3}
x^i_tx^j_t\eta^i_{x^j}+\frac{1}{2}x^i_tx^i_t(\xi^j_j-\xi^t_t)=0.
\end{align}
From the $p$ coefficient in \eqref{sb1} and the $x^i_tx^i_t$ coefficients in \eqref{sb3}, we get
\begin{align}\label{xit}
\xi^t(t)=(n+2)\lambda\,t+\gamma,\qquad \xi^j_j=n\lambda,\qquad \eta^i_{x^i}=\lambda,\qquad i=1,\dots,n,
\end{align}
where $\gamma$ and $\lambda$ are arbitrary constants.  The $x^i_tx^j_t,i\neq j$ coefficients in \eqref{sb3} give
\begin{align}\label{eta}
\eta^i(t,\vec x)=\lambda x^i+\omega^{ij}(t)x^j+b^i(t),\qquad i=1,\dots,n,
\end{align}
where $\omega^{ij}=-\omega^{ji}$.  We find from \eqref{sb2} that $N^t=x^ib^i_t$, $N^i=0$, and $b^i(t)=v^i t+\sigma^i$, where $v^i,\sigma^i$ are constants, and each $N^\mu$ is defined up to terms that do not influence $\X$.

Finally, for the requirement that $\xi^i_n=n\lambda$ in \eqref{xit}, we can substitute $\xi^i=\lambda a^i+\chi^i(\vec a)$.  We find that $\chi^i(\vec a)$ is an arbitrary divergenceless vector, or $\chi^i_i=0$.  In general, this means that $\chi^i=A^{ij}_j$ for some anti-symmetric matrix $A^{ij}(\vec a)=-A^{ji}(\vec a)$.




In summary, we find the variational point symmetry coefficients of $\X=\xi^\mu\partial_\mu+\eta^i\partial_{x^i}+\psi\partial_p$ and the fluxes admitted by action functional (\ref{action}) to be:
\begin{align}
\begin{split}
\xi^t(t)&=(n+2)\lambda t+\gamma,\\
\xi^i(\vec a)&=\lambda a^i+\chi^i(\vec a),\qquad i=1,\dots,n,\\
\eta^i(t,\vec x)&=\lambda x^i+\omega^{ij}x^j+v^it+\sigma^i,\qquad i=1,\dots,n,\\
\psi(p)&=-2(n+1)\lambda p,\\
N^t(\vec x)&=v^ix^i,\\
N^i&=0,\qquad i=1,\dots,n,
\end{split}
\end{align}
where $\lambda,\gamma,\omega^{ij}=-\omega^{ji},v^i,$ and $\sigma^i$ are arbitrary constants, and $\chi^i(\vec a)$ is an arbitrary vector function with zero divergence: $\chi^i_i=0$.

Since each constant is arbitrary, taking the coefficients of these constants gives independent symmetry generators.  The following symmetries are well known in the literature:
\begin{align}\label{syms_known}
\begin{split}
&\text{X}_\gamma=\partial_t,\\
&\text{X}_{\sigma^i}=\partial_{x^i},\qquad i=1,\dots,n,\\
&\text{X}_{v^i}=t\,\partial_{x^i},\qquad i=1,\dots,n,\\
&\text{X}_{\omega}=\omega^{ji}x^j\pd_{x^i},\\
&\text{X}_{\infty}=\chi^i(\vec a)\partial_{a^i}.
\end{split}
\end{align}
These symmetries respectively correspond to translations in time, translations in space, Galilean boosts, rotations in space, and volume-preserving relabelings of the particles (the subscript $\infty$ corresponds to ``infinite-dimensional").  

On the other hand, the following point symmetry of action (\ref{action}) and Euler-Lagrange equations \eqref{massL}-\eqref{eulerL} has yet to appear in the literature 
 for Lagrangian coordinates:
\begin{align}
\label{sym_scale}
&\text{X}_\lambda=(n+2)\,t\,\partial_t+a^i\partial_{a^i}+x^i\partial_{x^i}-2(n+1)\,p\,\partial_p.
\end{align}
This symmetry corresponds to a scaling symmetry of the action and Euler-Lagrange equations under $t\to\lambda^{n+2}t,a^i\to\lambda a^i,x^i\to\lambda x^i,$ and $p\to\lambda^{-2(n+1)}p$, $\lambda\neq 0$.  The factors of $n$ appear in the $t$ and $p$ variations due to the volume element $\di t\di^na$ in the action \eqref{action} having an $n$ number of $a$'s.  It is known to exist for the Eulerian formulation \eqref{mass}-\eqref{euler} as one part of a two-parameter scaling symmetry group \cite{Buchnev,Olver1982}, since the action of $\X$ on $\vec u=\vec x_t$ is $-(n+1)\vec u$, giving $\X_\lambda=(n+2)t\pd_t+x^i\pd_{x^i}-(n+1)u^i\pd_{u^i}-2(n+1)p\pd_p$ in Eulerian variables.  Symmetry operator $\X_\lambda$ is similar to one found by Kambe \cite{kambe} for an Eulerian variational formulation of irrotational compressible flow.  Note that \eqref{sym_scale} is also a symmetry of the action \eqref{S} for the classical variational formulation of incompressible flow.

\section{Symmetries and Conservation Laws}
In this section we find conservation laws of the Euler equations \eqref{massL}-\eqref{eulerL} corresponding to variational symmetries \eqref{syms_known}-\eqref{sym_scale}. 
In addition to the classical balance laws of kinetic energy, linear momentum, center of mass, angular momentum, and the statement of vorticity advection, we find a new conservation law for the Euler equations in Lagrangian coordinates corresponding to a scaling symmetry of the action functional. 

\subsection{Noether's Theorem}
We consider a first order Lagrangian $\mc L(x^\mu,x^\mu_\nu)$ with independent variables $a^\mu$ and dependent variables $x^\mu,\mu=t,1,\dots,n$.  The Euler-Lagrange equations are given by $E_\mu\mc L=0$, where $E_\mu$ is an Euler operator:
\begin{align}
E_\mu=\frac{\pd}{\pd x^\mu}-D_\nu\frac{\pd}{\pd x^\mu_\nu}+\dots,\qquad \mu=t,1,\dots,n.
\end{align}
Let $\X_\alpha=\X-\xi^\mu D_\mu=\alpha^\mu\pd_{x^\mu}+D_\nu\alpha^\mu\frac{\pd}{\pd x^\mu_\nu}+\dots$ be the canonical symmetry generator, where $\alpha^\mu=\eta^\mu-x^\mu_\nu\,\xi^\nu$ is the canonical infinitesimal.  The Noether operator identity \cite{Rosen}, see also \cite{Ibragimov1985}or \cite{Rosenhaus94}, connects $\X_\alpha$ to $E_\mu$:
\begin{align}\label{nid}
\X_\alpha=\alpha^\mu\,E_\mu+D_\mu R_{\alpha,\mu},
\end{align}
where $R_{\alpha,\mu}=\alpha^\nu\frac{\pd}{\pd x^\nu_\mu}+\dots$.  For variational symmetry $\X_\alpha$, applying \eqref{nid} to $\mc L$ in \eqref{invarX} gives
\begin{align}\label{ncon}
D_\mu\left(R_{\alpha,\mu}\,\mc L+\xi^\mu\,\mc L-N^\mu\right)=-\alpha^\mu\,E_\mu\mc L.
\end{align}
On solutions, ($E_\mu\mc L=0$ or $\doteq$), Eq. \eqref{ncon} associates to any variational symmetry $\X_\alpha$ a conservation law (Noether Theorem).  For $\mc L$ in \eqref{action}, we have $a^t=t,x^t=p$, so \eqref{ncon} takes the form
\begin{equation}\label{noether0}
D_t\left(\alpha^jx^j_t+\xi^t\mc L-N^t\right)+D_i\left(pJ\alpha^j\tilde{x}^{ij}+\xi^i\,\mc L-N^i\right)\,\dot{=}\,0.
\end{equation}
Of course, on solutions, we have $\mc L\doteq L=|\vec x_t|^2/2$, and $J\doteq 1$.  Thus, from \eqref{noether0}, we obtain:
\begin{equation}\label{noether}
D_t\left(\alpha^jx^j_t+\frac{1}{2}x^j_tx^j_t\,\xi^t-N^t\right)+D_i\left(p\alpha^j\tilde{x}^{ij}+\frac{1}{2}x^j_tx^j_t\,\xi^i-N^i\right)\,\dot{=}\,0.
\end{equation}


\subsection{Known conservation laws}
For the time translation symmetry $\text{X}_\gamma$ in (\ref{syms_known}), we have $\xi^t=1$ and $\eta^i=\xi^i=\psi=N^\mu=0$.  This gives $\alpha^i=-x^i_t$ as the canonical infinitesimal.  Substituting into the conservation law formula (\ref{noether}), we obtain the conservation form of energy balance:
\begin{equation}\label{c_energy}
D_t\left(\frac{1}{2}x^j_tx^j_t\right)+D_i\left(p\tilde{x}^{ij}x^j_t\right)\dot{=}\,0.
\end{equation}
This equation can be recast in Eulerian form.  Expanding the time derivative and using \eqref{xtt}, we have $$x^j_tx^j_{tt}=u^j(u^j_t+u^iu^j_{x^i})=D_t\left(\frac{1}{2}u^ju^j\right)+D_{x^i}\left(\frac{1}{2}u^iu^ju^j\right),$$
where \eqref{mass} was used. For the flux term, note the following operator identity: 
$$D_i=x^{j}_iD_{x_j}=D_{x_j}\cdot x^{j}_i-(\tilde x^{kj}D_kx^j_i)=D_{x_j}\cdot x^{j}_i-(D_k(\tilde x^{kj}x^j_i))=D_{x_j}\cdot x^{j}_i,$$
where we used \eqref{xiji} to conclude that $\tilde x^{kj}D_k=D_k\cdot \tilde x^{kj}$.  This gives:
\begin{equation}\nonumber
D_i\left(p\tilde{x}^{ij}x^j_t\right)=D_{x_k}\left(x^{k}_ip\tilde x^{ij}x^j_t\right)=D_{x^k}\left(px^k_t\right).
\end{equation}
Therefore, in vector notation, we obtain the usual kinetic energy balance law:
\begin{equation}
D_t\left(\frac{1}{2}|\vec u|^2\right)+\vec\nabla_x\cdot\left[\left(p+\frac{1}{2}|\vec u|^2\right)\vec u\right]\doteq 0.
\end{equation}

For the spatial translation symmetries $\text{X}_{\sigma^j}$, we have $\eta^{i,j}=\alpha^{i,j}=\delta^{ij}$ and $\xi^{\mu}=\psi=N^\mu=0$.  For each symmetry, substitution into conservation law formula (\ref{noether}) gives:
\begin{equation}\label{c_l_momentum}
D_t\left(x^j_t\right)+D_i\left(p\tilde{x}^{ij}\right)\,\dot{=}\,0,\hspace{4 mm} j=1,\dots,n.
\end{equation}
Expanding the derivatives and using $\tilde{x}^{ij}_i=0$ recovers momentum equation (\ref{eulerL}) in differential form. 

For the Galilean boosts $X_{v^j}$, we have $\eta^{i,j}=\alpha^{i,j}=t\delta^{ij}$, $N^{t,j}=x^j$, and $\xi^{\mu}=\psi=N^i=0$.  Substitution into \eqref{noether} gives:
\begin{align}
D_t\left(tx^j_t-x^j\right)+D_i\left(tp\tilde x^{ij}\right)\doteq 0,\qquad j=1,\dots,n,
\end{align}
which, upon integration over $\di^n a$, determines the motion of the $j-$th component of the flow's center of mass.  In Eulerian form:
\begin{align}
D_t\left(tu^j\right)+D_{x^i}\left(tu^ju^i+tp\delta^{ij}-x^ju^i\right)\doteq 0,\qquad j=1,\dots,n.
\end{align}
For rotational symmetry $\X_\omega$, we have $\eta^j=\alpha^j=\omega^{kj}x^k$, and $\xi^\mu=\psi=N^\mu=0$, so \eqref{noether} gives:
\begin{align}
D_t\left(\omega^{kj}x^kx^j_t\right)+D_i\left(p\tilde x^{ij}\omega^{kj}x^k\right)\doteq 0.
\end{align}
In Eulerian form:
\begin{align}
D_t\left(\omega^{ji}x^ju^i\right)+D_{x^i}\left(\omega^{kj}x^ku^ju^i+p\omega^{ji}x^j\right)\doteq 0.
\end{align}
If $n=2$, dividing by arbitrary constant $\omega^{12}=-\omega^{21}$ and letting $\ell=x^1u^2-x^2u^1$ gives
\begin{align}
D_t\ell+D_{x^1}\left(\ell u^1-px^2\right)+D_{x^2}\left(\ell u^2+px^1\right)\doteq 0.
\end{align}
If $n=3$, we have $\omega^{ij}=\omega^k\varepsilon^{kij}$.  For each arbitrary constant $\omega^k$, letting $\ell^k=\varepsilon^{kij}x^iu^j$ be the angular momentum components, we have
\begin{align}
D_t\ell^k+D_{x^i}\left(\ell^ku^i-p\varepsilon^{kij}x^j\right)\doteq 0,\qquad k=1,2,3,
\end{align}
which expresses the angular momentum conservation laws.

Symmetry $\X_\infty$ gives a relabeling transformation which maps the fluid labels $a^i$ to $a^{*i}=a^i+\epsilon\,\xi^i(\vec a)$ in such a way that preserves the volume element $\di^na$: $\xi^i_i=0$.  Newcomb \cite{Newcomb1967} used this symmetry transformation 
in conjunction with Noether's First theorem, in spite of the presence of an arbitrary function in its symmetry generator.  He then derived a conserved quantity using a fixed form of the arbitrary divergenceless functions $\xi^i(\vec a)$; in $n=3$ dimensions, this gave a representation formula for the fluid vorticity vector $\vec\omega\equiv\vec\nabla_x\times\vec u$, first obtained by Cauchy \cite{Cauchy}.  Note that the arbitrary functions $\xi^i(\vec a)$ depend on \textit{not all} independent variables; see \eqref{time}.  For arbitrary functions of not all independent variables, Rosenhaus \cite{Rosenhaus2002} showed that Noether's theorem does not yield an infinite set of conserved quantities when such arbitrary functions are involved: only one conserved quantity per arbitrary function of spatial variables, and a finite number of conserved quantities in the case when the arbitrary functions depend on time; see also \cite{Rosenhaus2003}.  Note that, according to Noether's Second Theorem, an infinite symmetry with arbitrary functions of \textit{all} independent variables would lead, instead of conserved quantities, to the conclusion that the equations of motion are underdetermined; see \cite{Rosenhaus2002}.

Using a similar approach to that in \cite{Rosenhaus2002}, we show that, instead of an infinite set of conserved quantities, this symmetry generates a finite number of first integrals of the Euler equations (Cauchy's invariants \cite{Cauchy}), recovering the representation formula for $\vec\omega$ when $n=3$.

In this case, we have $\alpha^j=-x^j_i\,\xi^i$, and $\eta^{i}=\xi^t=\psi=N^\mu=0$.  Substitution into (\ref{noether}) gives:
\begin{equation}\label{cont_inf}
D_t\left(x^j_tx^j_i\xi^i\right)+D_i\left[\left(p-\frac{1}{2}x^j_tx^j_t\right)\,\xi^i\right]\doteq 0.
\end{equation}
If we use that $\xi^i(\vec a)=A^{ij}_j(\vec a)$ for arbitrary antisymmetric matrix $A^{ij}=-A^{ji}$ and note that $$x^k_tx^k_iA^{ij}_j=D_j(x^k_tx^k_iA^{ij})-D_j(x^k_tx^k_i)A^{ij}=D_j(x^k_tx^k_iA^{ij})-x^k_{jt}x^k_iA^{ij}$$
(by antisymmetry, $x^k_{ij}A^{ij}=0$), we get a density without derivatives of $A^{ij}$:
$$
D_t\left(x^k_{jt}x^k_iA^{ij}\right)-D_i\left[D_t(x^k_tx^k_j)A^{ij}+\left(p-\frac{1}{2}x^k_tx^k_t\right)\,A^{ij}_j\right]\doteq 0.
$$
Let us now integrate over $\mathbb{R}^n$ and suppose that each $A^{ij}(\vec a)$ is an arbitrary function with compact support.  By Gauss' Theorem, the integral of the $D_i$ term vanishes, and we obtain
\begin{align}\label{inf}
\int_{\mathbb{R}^n}D_t\left(x^k_{jt}x^k_i\right)A^{ij}\di^na\doteq 0.
\end{align}
By the fundamental lemma of the calculus of variations \cite{Hilbert}, the arbitrariness of each $A^{ij}=-A^{ji}$ allows us to conclude that the integrand vanishes for all $\vec a$.  For example, if $n=2$, we can write $A^{12}(\vec a)=-A^{21}=-A(\vec a)$, so invoking the arbitrariness of $A$ gives:
\begin{align}\label{n2}
D_t\left(x^k_{1t}x^k_2-x^k_{2t}x^k_1\right)\doteq 0.
\end{align}
If $n=3$, we can write $A^{ij}(\vec a)=\varepsilon^{ijk}A^k(\vec a)$, so invoking the arbitrariness of each $A^k$ gives
\begin{align}\label{n3}
D_t\left(\varepsilon^{ijk}x^m_{jt}x^m_i\right)\doteq 0,\qquad k=1,2,3.
\end{align}
This verifies that, rather than an infinite set of conserved quantities, \eqref{cont_inf} leads to a finite number, which in this case take the form of first integrals of the equations of motion.  We evaluate these integrals and show how they correspond to the vorticity $\vec\omega$.

If $n=2$, we use that $x^k_{j t}=u^k_j=u^k_{x^\ell}x^\ell_j$ to rewrite \eqref{n2} as
$$
u^k_{x^\ell}\left(x^k_2x^\ell_1-x^k_1x^\ell_2\right)=u^2_{x^1}\left(x^2_2x^1_1-x^2_1x^1_2\right)+u^1_{x^2}\left(x^1_2x^2_1-x^1_1x^2_2\right)=c(\vec a),
$$
where $c(\vec a)$ does not depend on $t$.  But in two dimensions, the vorticity is perpendicular to the plane: $\vec\omega=\omega\hat z$, with $\omega=u^2_{x^1}-u^1_{x^2}$.  We conclude that the left hand side is $\omega\det(x^i_j)=\omega$ by \eqref{massL}, and $c=\omega$.  Thus,
\begin{align}
\omega(\vec a,t)=\omega(\vec a,0),
\end{align}
which shows that the vorticity does not depend on time and is transported with the particles by the velocity flow.

Let us now consider the $n=3$ case \eqref{n3}.  We obtain the following first integral of the equations of motion (\ref{eulerL}):
\begin{equation}\label{v_int}
\varepsilon^{ijk}x^m_{jt}x^m_k=c^i(\vec a),
\end{equation}
where $c^i=\varepsilon^{ijk}x^m_{jt}x^m_k|_{t=0}$.

We now solve (\ref{v_int}) for the vorticity $\vec\omega$.  We have $x^m_{jt}=u^m_j=u^m_{x^q}x^q_j$, so multiplying \eqref{v_int} by $x^p_i$ gives
$$
\varepsilon^{ijk}x^p_ix^q_jx^m_ku^m_{x^q}=c^ix^p_i.
$$
By a well known formula, $\varepsilon^{ijk}x^p_ix^q_jx^m_k=\varepsilon^{pqm}\det(x^i_j)$, so using \eqref{massL} gives:
$$
\varepsilon^{pqm}u^m_{x^q}=c^ix^p_i.
$$
But $\varepsilon^{pqm}u^m_{x^q}=\omega^p$ is the $p-$th component of the vorticity vector.  Since $x^p_i|_{t=0}=\delta^{pi}$, we see that $c^i(\vec a)=\omega^i(\vec a,0)$ is the initial vorticity.  In vector notation:
\begin{equation}\label{c_vort}
\vec\omega(\vec a,t)=\vec\omega(\vec a,0)\cdot\vec\nabla_a\,\vec x(\vec a,t).
\end{equation}
This representation formula states that the vorticity at any time has been advected from the initial vorticity by the Jacobian matrix of Lagrangian map $\vec x$.  Depending on the flow gradient $\vec\nabla_a\vec x$, this may lead to enhancement of $\vec\omega$ (vortex stretching). 

\subsection{Scale-Invariance Conservation Law}
Consider now a conservation law corresponding to scaling symmetry (\ref{sym_scale}) of action functional (\ref{action}).  Here, $\eta^i=x^i$, $\xi^i=a^i$, $\xi^t=(n+2)t$, $\psi=-2(n+1)p$, and $N^\mu=0$.  These infinitesimals produce canonical infinitesimal $\alpha^j=x^j-a^kx^j_k-(n+2)tx^j_t$, so substitution into Noether conservation law (\ref{noether}) gives:
\begin{align}\label{c_adv}
\begin{split}
&D_t\left[\left(x^j-a^kx^j_k-\frac{n+2}{2}tx^j_t\right)x^j_t\right]\\
&+D_i\left[p\tilde{x}^{ij}\left(x^j-(n+2)tx^j_t\right)+\left(\frac{1}{2}x^j_tx^j_t-p\right)a^i\right]\doteq 0.
\end{split}
\end{align}

To the author's knowledge, this conservation law has yet to appear in the literature, even in \cite{Caviglia1989}, where the authors performed an exhaustive classification of the set of conservation laws of the incompressible Euler equations.  Kambe \cite{kambe} obtained a similar conservation law for the distinct case of irrotational compressible flow using Noether's theorem and a scaling symmetry.  The main difference between his and our conserved densities seems to be the term $-a^kx^j_kx^j_t$, which has no counterpart in his Eulerian formulation.

Let
\begin{align}\label{int}
I\equiv \int \left(x^j-a^kx^j_k-\frac{n+2}{2}\,t\,x^j_t\right)\,x^j_t\,\di^na,\qquad t\ge 0.
\end{align}
be the conserved (constant) integral of conservation law \eqref{c_adv}.  For conservation law \eqref{c_energy}, we let
\begin{align}\label{E}
E\equiv \int\frac{1}{2}x^j_tx^j_t\di^n a,\qquad t\ge 0.
\end{align}
be its conserved integral.  Here, $E$ corresponds to the total (kinetic) energy of the fluid.  

We show that conserved integral $I$ relates energy $E$ to a new integral quantity.  Equality \eqref{int} must hold at $t=0$.  At this time, we have $\vec x(\vec a,0)=\vec a$, so $x^j=a^j$, and $x^j_k=\delta^{jk}$.  Therefore, $x^j-a^kx^j_k=0$, so constant $I$ in \eqref{int} must be zero at $t=0$, hence for all $t\ge 0$.  This gives the relation
\begin{align}\label{int1}
\int\left(x^j-a^kx^j_k-\frac{n+2}{2}\,t\,x^j_t\right)x^j_t\di^{n}a=0,\qquad t\ge 0.
\end{align}
Observe that the last term in \eqref{int1} can be written as
$$
-\frac{(n+2)t}{2}\int x^j_tx^j_t\di^na=-(n+2)\,t\,E,
$$
where $E$ is as in \eqref{E}.  If we solve \eqref{int1} for $E$ and recall that $\vec u=\vec x_t$, we find that
\begin{align}\label{Ea}
E=\frac{1}{(n+2)\,t}\int \left(\vec x-\vec a\cdot\vec\nabla_a\,\vec x\right)\cdot\vec u\,\di^na,\qquad t>0,
\end{align}
which represents $E$ as a \textit{different} integral.  

Although $E=\int\frac{1}{2}|\vec u|^2\di^n a=\int\frac{1}{2}|\vec u|^2\di^n x$ is a well defined Eulerian quantity (from \eqref{massL}, $\di^na=\di^n x$), the integrand in \eqref{Ea} has no Eulerian counterpart, due to the term $-\vec a\cdot\vec\nabla_a\vec x\cdot\vec u$.  Because it is not possible to convert to Eulerian form, this integrand is exclusively defined in Lagrangian coordinates.  It is interesting how an Eulerian quantity can also have a strictly Lagrangian representation.

We note that \eqref{Ea} has a definite limit as $t\to 0$.  Let $\vec u_0(\vec a)=\vec u(\vec a,t)$ be the initial velocity.  By \eqref{t0}
\begin{align*}
\frac{1}{t}\left[\vec x-\vec a\cdot\vec\nabla_a\vec x\right]\cdot\vec u&=\frac{1}{t}\left[\vec a+t\vec u_0-\vec a\cdot\vec\nabla_a(\vec a+t\vec u_0)+O(t^2)\right]\cdot\left[\vec u_0+O(t)\right]\\
&=\left(\vec u_0-\vec a\cdot\vec\nabla_a\vec u_0\right)\cdot\vec u_0+O(t),
\end{align*}
so taking the limit as $t\to 0$ in \eqref{Ea} gives
\begin{align}\label{E1}
E=\frac{1}{n+2}\int \left(\vec u_0-\vec a\cdot\vec\nabla_a\vec u_0\right)\cdot\vec u_0\di^na.
\end{align}
On a related note, since \eqref{E} holds at $t=0$, we have $E=\frac{1}{2}\int|\vec u_0|^2\di^na$.  Equating this to the right hand side of \eqref{E1} shows that equality holds provided that 
\begin{align}\label{necc}
\int \vec\nabla_a\cdot\left(\frac{1}{2}|\vec u_0|^2\,\,\vec a\right)\di^na=0.
\end{align}
is satisfied.  This is a necessary condition for $I$ to be conserved.  In general, it will be satisfied when $E$ is finite.

We now give a physical interpretation for what integral formula \eqref{Ea} suggests.  Since $E$ is conserved, the right hand side of \eqref{Ea} must be constant for all time $t>0$.  Therefore, the integral must be proportional to $t$, which means it becomes large as $t\to\infty$.  But by Schwarz's inequality,
$$
\left|\int\left(\vec x-\vec a\cdot\vec\nabla_a\,\vec x\right)\cdot\vec u\,\di^na\right|\le\left[\int\left|\vec x-\vec a\cdot\vec\nabla_a\,\vec x\right|^2\di^n a\right]^{1/2}(2E)^{1/2},
$$
so the only part of the integrand in \eqref{E} that can grow large is $\vec x-\vec a\cdot\vec\nabla_x\vec x$.  Specifically,
\begin{align}\label{est}
\int\left|\vec x-\vec a\cdot\vec\nabla_a\,\vec x\right|^2\di^n a\,\,\ge\,\,\frac{(n+2)^2}{2}\,\,E\,t^2,\qquad t\ge 0.
\end{align}
Since $\vec a\cdot\vec\nabla_a=|\vec a|\frac{\pd}{\pd|\vec a|}$ is a radial derivative, we interpret $\vec x-\vec a\cdot\vec\nabla_a\vec x$ as the \textit{radial deformation} of the particle trajectory $\vec x(\vec a,t)$.   The left hand side of \eqref{est} then measures the total radial deformation of the fluid at time $t$.  
We illustrate this with an example.

Let us consider a well known \cite{Majda} exact solution of the Eulerian formulation \eqref{mass}-\eqref{euler} for $n=2$:
\begin{align}\label{esol}
\begin{split}
u^1(\vec x,t)&=-\Omega(|x|)\,\,x^2,\\
u^2(\vec x,t)&=\Omega(|x|)\,\,x^1,\\
p(\vec x,t)&=p_0+\int_0^{|x|}r\,\Omega(r)^2\di r,
\end{split}
\end{align}
where $\Omega(|x|)$ is an arbitrary smooth radial function, and $p_0$ is a constant.  Energy \eqref{E} takes the form 
\begin{align}\label{Epolar}
E=\pi\int_0^\infty r^3\Omega(r)^2\di r.  
\end{align}
Solution \eqref{esol} describes an ideal fluid with an angular velocity that depends only upon distance from the origin (a radial eddy).  There is no radial velocity, so particles transported by the flow retain their original distance to the origin.

To find the particle trajectories in Lagrangian coordinates, we integrate the system $\vec x_t=\vec u(\vec x,t)$ and impose the initial condition $\vec x(\vec a,0)=\vec a$.  Letting $\theta(\vec a)=\arctan(a^2/a^1)$ be the angle in the Cartesian plane, we obtain
\begin{align}\label{lsol}
\begin{split}
x^1(\vec a,t)&=|\vec a|\,\cos\left(\Omega(|\vec a|)t+\theta(\vec a)\right),\\
x^2(\vec a,t)&=|\vec a|\,\sin\left(\Omega(|\vec a|)t+\theta(\vec a)\right).
\end{split}
\end{align}
Observe that $\Omega(|\vec a|)$ is the \textit{angular velocity} of the particle given by initial position $\vec a$.  For this solution, a direct calculation shows that the radial deformation is as follows:
\begin{align}\label{def}
\begin{split}
x^1-\vec a\cdot\vec\nabla_a x^1&=t\,|\vec a|\,\Omega'(|\vec a|)\,\,x^2,\\
x^2-\vec a\cdot\vec\nabla_a x^2&=-t\,|\vec a|\,\Omega'(|\vec a|)\,\,x^1.
\end{split}
\end{align}
Comparing with \eqref{esol}, we see that $\vec x-\vec a\cdot\vec\nabla_a\vec x=-t|\vec a|\Omega'(|\vec a|)\,\,\vec u\,\,/\Omega(|\vec a|)$.

If $\Omega'=0$, or if $\Omega(|\vec a|)$ is a constant angular velocity, then \eqref{def} shows that the radial deformation is zero.  Indeed, in this case, particles of large radius $|\vec a|\to\infty$ revolve around the origin as quickly as those near the origin, $|\vec a|\to 0$.  This represents rigid-body motion, for which no fluid deformation occurs.  On the other hand, if $\Omega'\neq 0$, then \eqref{def} shows that radial deformation results.  Indeed, since the angular velocity $\Omega$ varies with radius $|\vec a|$, near particles move differently from far particles, so the motion is not rigid.  In particular, if $\Omega(0)>0$ but $\Omega\to 0$ as $|\vec a|\to\infty$, then the angular velocity can diminish with distance from the origin.  In general, this is case for finite energy solutions.

From \eqref{def}, there is no fluid deformation for small times $t\to 0$, while the particles are still at their initial positions.  On the other hand, as $t$ increases, the fluid deformation increases without bound.  Initially close particles of slightly different radii $|\vec a|$ and $|\vec a|+\delta$ become separated from each other due to the disparity in their angular velocities $\Omega(|\vec a|)$ and $\Omega(|\vec a|)+\delta \Omega'(|\vec a|)$.  The deformation increases in \eqref{est} with energy $E$ because larger $E$ in \eqref{Epolar} means larger $\Omega$, or larger angular velocities.  Since $\Omega\to 0$ as $|\vec a|\to\infty$, larger $\Omega$ near $|\vec a|=0$ means the angles of near and far particles separate more quickly.

In general, if $\vec x-\vec a\cdot\vec\nabla_a\vec x=0$, then trajectories are homogeneous in $|\vec a|$, or $\vec x=|\vec a|\,\,\vec y(\vec a/|\vec a|,t)$, where $\vec y$ depends only on direction $\vec a/|\vec a|$ and time $t$.  For example, $\vec x(\vec a,t)=\vec a$ (constant flow with zero velocity), or more generally $\vec x(\vec a,t)=M(t)\,\vec a$ for some $n\times n$ matrix $M(t)$ with determinant equal to $1$ (rigid-body motion).  Such trajectories may exist for some $\vec a$ in $\re^n$, but not all, if $Et^2>0$.  Moreover, since the right hand side of \eqref{est} increases with time, this inequality reflects that such trajectories become less important to the overall behavior of the flow as time increases.



Surprisingly, \eqref{est} invalidates the existence of smooth, time-periodic solutions of nonzero, finite energy.  Indeed, if $\vec x(\vec a,T)\equiv\vec x(\vec a,0)=\vec a$ for some $T>0$ and all $\vec a$, then the integral in \eqref{est} vanishes, implying the right side is zero.  Since $T>0$, we conclude that $E=0$.  Note that \eqref{lsol} is \textit{not} a periodic solution if it has finite energy (despite being composed of periodic trajectories), since the angular velocity changes continuously with $\vec a$ (if $\Omega=$ constant, such that $\vec x$ is periodic, then $E=\infty$ from \eqref{Epolar}).  Note that there exist time-periodic solutions with finite energy for the Navier-Stokes equations (Eulerian formulation) with a periodic forcing term \cite{Hu,Silvestre}.  For compressible flow, the existence of time-periodic solutions is an open problem \cite{Temple}.

More is true: we cannot have $\vec x(\vec a,\tau+T)\equiv \vec x(\vec a,\tau)$ for any $\tau\ge 0$ or $T>0$ (and all $\vec a$).  To show this, we let $\tau\ge 0,T>0,$ and change variables $(t,\vec a)\to(s,\vec b)$, $\vec x(\vec a,t)=\vec y(\vec b,s)$, such that $\vec x(\vec a,\tau)$ becomes the initial position of the particle trajectories: $s=t-\tau, \vec b=\vec x(\vec a,\tau)$, and $\vec y(\vec b,0)=\vec b$.  But the equations of motion \eqref{massL}-\eqref{eulerL} are invariant under time translations and transformations that preserve volume element $\di^na$ (cf. $\X_\gamma$ and $\X_\infty$ in \eqref{syms_known}), of which, by \eqref{massL}, $\vec a\to\vec x(\vec a,\tau)$ is a member.  This means $\vec y(\vec a,t)$ is \textit{also} a solution, so we must have $\vec y(\vec a,T)\neq \vec y(\vec a,0)$ for some $\vec a$.  Since $\vec a\to \vec x(\vec a,\tau)$ is a bijective map (for all sufficiently small times; see Lemma 4.4 and Theorem 4.2 in \cite{Majda}), this means that $\vec y(\vec x(\vec a_1,\tau),T)\neq \vec y(\vec x(\vec a_1,\tau),0)$ for some $\vec a_1$.  However, by definition of $\vec y$, this means that $\vec x(\vec a_1,\tau+T)\neq \vec x(\vec a_1,\tau)$, which is what we wanted to show.   
  
For this reason, we find that radial deformation of an ideal fluid flow with finite energy is not time-reversible, such that the bulk flow cannot return to any of its previous states.


\section{Conclusion}
We obtained the following results for the incompressible Euler equations in Lagrangian coordinates:

We discussed an action functional (\ref{action}) for the incompressible Euler equations.  Our formulation is a modification of that in \cite{Caviglia1987} used for compressible flows 
and based off an indirect classical formulation \eqref{S}.  Using this action functional, we clarified the relationship between symmetries and conservation laws of the incompressible Euler equations; we showed that point symmetries of the action result via Noether's theorem in conservation laws of energy, momentum, center of mass, angular momentum, and the statement of vorticity advection.  

It is known \cite{Olver} that, for variational systems, there is a one-to-one correspondence between variational symmetries and local conservation laws; therefore, all local conservation laws of a given order are generated by corresponding variational symmetries of this order.  In this paper, we have found all lower order conservation laws corresponding to point symmetries.  However, the conservation of vorticity, helicity \cite{Olver1982}, and generalized linear momentum \cite{Caviglia1989} indicate the possible existence of generalized (higher) variational symmetries of the Euler equations in Lagrangian coordinates whose coefficients $\xi^\mu,\eta^i,\psi$ depend on derivatives of $x^i,p$.

We found a new point symmetry (\ref{sym_scale}) of action (\ref{action}) corresponding to a scaling invariance of the action functional and equations of motion.  This symmetry is similar to one found in \cite{kambe} for compressible potential flows.

Using this symmetry, we constructed a new conservation law (\ref{c_adv}) in Lagrangian coordinates.  The conserved integral \eqref{int} relates the total kinetic energy \eqref{E} of the fluid, an Eulerian quantity, to a new integral quantity \eqref{Ea} defined exclusively in Lagrangian coordinates.  This relationship implies an inequality \eqref{est} which shows that the total radial deformation of the fluid increases with both time and total kinetic energy.  The deformation was shown to be time-irreversible, proving the non-existence of time-periodic solutions with nonzero, finite energy.  


\section*{Acknowledgments}
The author is grateful to Tucker Hartland for his contributions and to Prof. V. Rosenhaus for stimulating discussions and useful suggestions.

\section*{References}

\bibliography{bibfile}

\end{document}